# Computer-mediated communication in adults with high-functioning Autism Spectrum Conditions

Authors:
Christine P.D.M. van der Aa
Tilburg University, Tilburg, The Netherlands

Monique M.H. Pollmann
Tilburg University, Tilburg, The Netherlands

Aske Plaat
Leiden University, Leiden, The Netherlands
(Corresponding author. Email: aske.plaat@gmail.com)

Rutger Jan van der Gaag
Radboud University, Nijmegen, The Netherlands

**Abstract**

It has been suggested that people with Autism Spectrum Conditions (ASC) are attracted to computer-mediated communication (CMC). In this study, several open questions regarding CMC use in people with ASC which are investigated. We compare CMC use in adults with high-functioning ASC (N = 113) and a control group (N = 72). We find that people with ASC (1) spend more time on CMC than controls, (2) are more positive about CMC, (3) report relatively high levels of online social life satisfaction, and that (4) CMC use is negatively related to satisfaction with life for people with ASC. Our results indicate that the ASC subjects in this study use CMC at least as enthusiastically as controls, and are proficient and successful in its use.

Computer-Mediated Communication in Adults with High-Functioning Autism Spectrum Conditions

Despite the interest in Internet-based social media in the popular and scientific literature, relatively little attention has been paid to the impact of computer-mediated communication (CMC) on people with Autism Spectrum Conditions (ASC). The popularity of autism-related websites and mailing lists suggests high online activity by people with ASC[1] (Davidson, 2008). However, people with ASC communicate less and less well than people without ASC (American Psychiatric Association, 2000, 2010; Hengeveld, Van Londen, & Van der Gaag, 2008). The question is then, to what extent do people with ASC make use of online communication channels when they are online, and how do they perceive this type of communication. Another question is whether CMC helps people with ASC to have a richer social life. In the current study, we investigate these questions by comparing uses and outcomes of CMC between people with and without ASC.

**Computer-mediated communication**

Computer-mediated communication is relatively consistent, predictable, and uni-modal; most CMC is text-based, takes place in a structured environment, is frequently asynchronous (giving users more time to process the information) and has fewer distracting signals (it should be noted that CMC is not by definition asynchronous as there are also channels with synchronous communication, but these constitute only a small part of CMC). Also, CMC often provides spatial and temporal distance between communication partners, and allows working at one's own convenience and pace, which fits the needs of people with ASC well. Several studies indicate that there is a special attraction to the Internet and computer-based tools for people with ASC, as we will cover in the discussion section (e.g., Cheng, Kimberly, & Orlich, 2002; Grynszpan, Martin, & Nadel, 2008; Ramdoss et al., 2011;

Finkenauer, Pollmann, Begeer, & Kerkhof, 2012). The conclusion of these studies is that the text-based nature of CMC affords a reduced-cues method of communication, greatly reducing the sensory overload that many ASCs experience, and thereby leading to improved communication.

Given their attraction to computers it seems likely that people with ASC also make frequent use of CMC. Indeed, a recent study on social media use among adults with ASC showed that 80% use social media and spend on average 3 hours a day using them (Mazurek, 2013). Additionally, a survey among 138 people with ASC showed that text-based, asynchronous communication channels were preferred to traditional forms of communication and that people with ASC report a high level of internet use in general (Benford, 2008). So, on the one hand, people with ASC tend to communicate less and on the other hand CMC seems to be an attractive way of communicating for them. However, these studies do not compare CMC use of people with ASC with a control group, so we do not know whether people with ASC use CMC more than a comparable control group. In the current study we compare CMC use of people with and without ASC. We hypothesize that CMC is used at least as much by people with ASC as by controls.

**Characteristics of CMC**

In addition to the questions whether people with ASC make more use of CMC than controls, it is also interesting to investigate whether people with ASC value different aspects of CMC than controls. There are some studies that mention which aspects of CMC are liked by people with ASC. For example, Benford (Benford, 2008; Benford & Standen, 2009) interviewed people with ASC and found that online communication provides them with a sense of liberation, afforded by specific characteristics of CMC. The first of these characteristics is control, both over the timing (communicating at a self-selected time) and pacing (immediacy of response) of a conversation, and over the way one can present oneself.

Another main point was the clarity of written text; more structured, and with less social chit-chat than in real life. The absence of non-verbal cues was important for diminishing the stress brought about by real-life conversations. Disadvantages mentioned included the risk of disclosing too much personal information, and not knowing whether the communication partner can be trusted. Burke, Kraut, and Williams (2010) found similar themes, describing the attractiveness of features such as CMC's slower pace and the absence of non-verbal signals and of the need for making eye contact. Furthermore, the possibilities to find like-minded people and to use predefined emoticons were named as benefits. Problems encountered by people with ASC included knowing whom to trust, and how to maintain a relationship. A study by Davidson (2008) suggests that the emergence of an autistic culture online is supported greatly by special features of CMC such as its slower pace, the ability to communicate with like-minded people, and the absence of the demands of physical co-presence.

The overall picture emerging from these studies is that there are many aspects of CMC that are advantageous for people with ASC. However, again, a common weakness of these studies is that they have been done without a control group. This makes it difficult to judge whether the value mentioned so often is really different from the value a general population attributes to online communication. Also, it remains unclear which features are specifically useful for ASCs, as most of the features that seem advantageous have also been named as such in a more general context. In the current study, we therefore compare the perceptions of CMC of people with ASC and a control group. We hypothesize that people with ASC will value different aspects of CMC than controls.

**CMC use and well-being**

Although some early studies on the link between CMC and well-being in the general population suggested that CMC can have a negative impact on people's social life (Kraut,

Patterson, Lundmark, Kiesler, Mukophadhyay, & Scherlis, 1998; Nie & Erbring, 2000), more recent studies link CMC use to various positive social outcomes (Amichai-Hamburger & Furnham, 2007). For example, Valkenburg & Peter (2007) find a positive relationship between instant messaging, and time spent with existing friends and the quality of those friendships. Additionally, researchers have noted how the internet offers an additional set of tools for getting acquainted with people, and maintaining these contacts (Ellison, Steinfield, & Lampe, 2007; Orr, et al., 2009; Steinfield, Ellison, & Lampe, 2008). Finally, some authors mention how the internet functions as a means for acquiring and improving one's social skills (Amichai-Hamburger & Furnham, 2007). For instance, games are seen as an ideal platform for practicing these skills by providing a topic to talk about and the option to hide behind an avatar (Ducheneaut & Moore, 2005). Whether these advantages hold in the same way for people with ASC is not yet fully clear, as there are few systematic, controlled studies relating CMC use of people with ASC to life outcome variables, with some notable exceptions, as described below.

Davidson (2008) and Mitchell (2003) studied autistic culture online and found that the ability to have meaningful communication without the need to respond immediately, and the slower pacing of CMC in general, could alleviate the stress that many ASCs encounter during real-life encounters. Benford and Standen (2009) surveyed people with Asperger's Syndrome and high-functioning autism about their experiences and perceptions of CMC. Their subjects report that CMC has helped them expand their social networks and get more social support, decreasing feelings of loneliness. However, a recent study on social media use in adults with ASC found no link between frequency of use and feelings of loneliness (Mazurek, 2013).

The only study to use a control group is a study of word usage in blogs by Newton, Kramer, and McIntosh (2009). Interestingly, word usage was found to be almost identical in the two groups, except for the use of social words, which was more variable in ASCs than in controls. Their conclusion was that online there might be little difference in communication between the two groups, and that social-communicative deficits of ASCs could be induced by the proximal setting in which traditional social contact takes place, rather than being an impairment per se. Newton and colleagues suggest that in a more distal setting, as provided by CMC, the manifestation of these deficits may be diminished or even absent. This view is in contrast to the common view that people with ASC lack interest in contact with others, as evidenced by some of the diagnostic and symptomatic criteria (APA, 2000, 2010). In this view people with autism prefer numbers and things over people. Contrary to this view, the findings discussed above suggest that people with ASC may not lack interest in social contact, but lack social skills, as required in the everyday, proximal setting of face-to-face conversations. These studies imply that, contrary to the stereotype, high-functioning ASCs are interested in having relationships with other people (see, for example, Benford & Standen, 2009; Burke, et al., 2010; Davidson, 2008; Newton, et al., 2009). Given that CMC offers them the opportunity to have a social life which is easier to manage, we hypothesize that people with ASC can develop a satisfactory online social life.

    What is more, CMC may be able to improve general well-being. ASC is related to higher levels of loneliness (Jobe & Williams-White, 2007), but having a good support network is positively related to quality of life in people with ASC (Renty & Roeyers, 2006). If the online social life can function as a support network, CMC should be positively related to indicators of well-being in people with ASC. We will therefore not only test whether people with ASC are satisfied with their online social life, but also whether CMC is related to more general indicators of well-being. We hypothesize that CMC can improve general well-being in people with ASC.

## Method

**Participants**

Since this study focuses on people that use the Internet for computer-mediated communication, participant requests were primarily distributed via online channels. For the ASC group, a request to participate was posted on the LinkedIn discussion group "Autisme Ten Top" (Autism Par Excellence), the websites and newsletters from autism organizations NVA (Nederlandse Vereniging voor Autisme or Dutch Association for Autism, originally a parent organization) and PAS ("Personen uit het Autisme Spectrum" or Persons from the Autism Spectrum, an organization for and by adults on the spectrum, with normal to high intelligence) and the autism discussion group "Autsider." In addition, some recruiting was done through one of the authors' personal network, and several health care organizations were asked to cooperate in finding candidates, for instance during an open house (April 2010) and through e-mail and telephone requests.

For the non-ASC group, recruitment was more complicated, because there is no online platform for the 'general' individual. We therefore asked students and acquaintances to invite people in their network to participate, and if possible to forward the request to others in their social network. We deliberately did not invite people from our own network to participate, but made sure that he invitation was sent at least one degree further to reduce the selection bias of middle-aged, highly educated individuals and collect a sample that is as similar as possible to the ASC group. To this end, we also explicitly asked to distribute the request over younger and older people, of different educational levels, and throughout the country. Another prerequisite for participation was that individuals were familiar with online communication, again to ensure similarity between the ASC and the control group.

It should be noted that as our subjects are people that use online communication channels, our findings do not generalize to people that do not use CMC, or to non-high-functioning ASCs, as we will discuss in the discussion section.

We received data from 203 individuals, but in one case, the caregiver instead of the individual with ASC themselves completed the survey and two respondents did not meet the requirement that respondents needed to be at least 18 years old, so these responses were not included. Due to technical problems 20 questionnaires were incomplete and are not included in each analysis reported below. Information on gender was available from 183 respondents, with 90 men and 93 women in the sample. Respondents were asked whether they had a diagnosis in the autism spectrum. Respondents were considered to belong to the ASC group if they answered yes to this question (108 individuals), or if they self-identified as ASC in their comments (5 individuals). We included the self-identified ASC in the ASC group, since these individuals indicated that they are certain that they classify as ASC.[2] Based on these criteria, the ASC group counted 113 respondents, and the control group 72 respondents. Table 1 provides demographic statistics on the subjects.

**[insert Table 1 here]**

The recruitment method resulted in a higher-than-average level of education in both the control and the ASC group. We included a question about the completed and the uncompleted level of education, which indicates subjects that are still enrolled at a school, or have dropped out. Since ASCs tend to have more difficulties with the transition from school to college (VanBergeijk, Klin, & Volkmar, 2008), the uncompleted level of education may give a better indication of their intellectual capabilities than the completed level of education. Most respondents had the Dutch nationality; some were Belgian. No information about ethnic background, socio-economic status or sexual orientation was available.

When we compare the ASC group to the control group we find that the ASC group in the sample: has almost the same age (40 years), contains more men (55.9% vs. 38.9%), is more often single (59.6% vs. 27.8%), is living independently less often (84.7% vs. 95.5%), is more often unemployed or living on a disability allowance (53.2% vs. 7.0%), and both the completed and uncompleted level of education is lower. These results are in line with what we expected of a group of high-functioning ASCs versus a control group recruited via students and acquaintances of researchers, and we have no evidence that the differences are of a quality that they can explain differences in perceptions of CMC.

**Measures and Procedure**

The survey was held from May 25, 2010 to June 25, 2010. The survey was conducted in Dutch, using an online survey tool. Two € 15 book vouchers were raffled off among all respondents who had completed the survey and had entered their e-mail address for this goal. All persons gave their informed consent prior to their participation in the study.

**Internet and CMC use.** *Internet use* and *CMC use* were measured in hours per week, calculated from the number of days per week (1-7) or per month (1-3), and hours per day (0 - 10 hrs or more, in half hour blocks) that respondents spent online or on CMC. This way of measuring internet and CMC usage was based on the method used by Valkenburg & Peter (2007), except here no distinction was made between weekdays and weekends. To get a more specific picture of the channels used, we asked people to indicate for nine *different channels* how often they used this channel. Answers were scored on a 6-point scale, ranging from 5 (more than 2 times per day) to 0 (less than once per month). In addition, respondents were asked if they had *found friends or acquaintances* through CMC that they would not have known otherwise. If they indicated that they did, they were further asked to indicate *how many friends and acquaintances* they found through CMC. Finally, we measured *appreciation of CMC* with 5 items regarding the value a respondent places on different aspects of online communication. The items were scored on a 7-point Likert scale, ranging from 1 (totally disagree) to 7 (totally agree) and had a good internal consistency, α = .84.

**Characteristics of CMC.** We assessed the perceived advantages and disadvantages of CMC with two types of questions. We first asked in two open questions to *list the advantages and disadvantages* of CMC as perceived by the respondent. The answers to these open-ended questions were categorized, based both on the answers themselves, and on the themes found in the literature. These fine-grained categories were collapsed into major clusters, 9 for advantages, and 8 for disadvantages.

Second, we presented respondents with a list of statements about CMC, which was created based on the advantages that had been reported in previous studies (e.g. Benford & Standen, 2009; Burke, et al., 2010). These included: pacing of the conversation, absence of non-verbal communication, anonymity, ability to find like-minded people, etc. For each characteristic, respondents were asked to indicate on a visual analog scale (Ahearn, 1997) whether they saw this *characteristic as a disadvantage or an advantage*. The starting point of the slider was in the middle of the scale and respondents could move it to the left to indicate a disadvantage (down to 1) or move it to the right to indicate an advantage (up to 100). As described in the results section, we performed factor analyses to identify underlying clusters in these characteristics of CMC. Based on these analyses we created three subscales: *Time independence* with 15 items (α = .89), *No co-presence* with 8 items (α = .87), and *Relative ease to express oneself* with 5 items (α = .76) (see Table 5 for the complete list).

**Well-being scales**. To investigate whether more use of CMC can have positive consequences in people's life, we asked respondents to indicate their satisfaction for different aspects of their life. The aspects ranged from concrete to abstract; with the most concrete being *satisfaction with one's online social life* (α = .82), then *satisfaction with one's social life* (α = .95), then *satisfaction with life* (α = .94), and finally their general *loneliness* (α = .84).

The satisfaction scales each consisted of 5 items measured on a 7-point Likert scale following Diener's satisfaction with life scale (Diener, Emmons, Larsen, & Griffin, 1985). Scores could range from 1 (totally disagree) to 7 (totally agree). The loneliness scale consisted of 6 items based on the UCLA Loneliness Scale (Russell, Peplau, & Cutrona, 1980), for example 'Do you ever feel lonely?'. These items were measured on a 5-point Likert scale with answer categories ranging from 5 (never) to 1 (always). Answers were scored such that higher scores represent less loneliness.

**Autism Spectrum Quotient (AQ).** As an additional check for the distinction between people with and without ASD, participants were asked to fill out the AQ, a self-report questionnaire, originally developed by Baron-Cohen, Wheelwright, Skinner, Martin, and Clubley (2001). The translation used here was the Dutch Autism Spectrum Quotient (Hoekstra, Bartels, Cath, & Boomsma, 2008). The reliability of the scale was good $\alpha = .96$, and the ASC group scored significantly higher ($M = 34.68$, $SD = 7.88$) than the controls ($M = 13.59$, $SD = 8.10$), $t(179) = 17.29$, $p < .001$, $r^2 = .67$, see also Table 1. Three out of the 72 controls scored an AQ above the commonly suggested threshold of 32, but we decided to still treat them as controls and not re-assign them to the ASC group as they were recruited as controls. The full questionnaire that was used is available online (Blinded) and in (Blinded).

## Results

**Internet and CMC use**

Our first aim was to test whether people with and without ASC use the Internet and CMC differently. To prevent an inflation of the type I error we performed a MANOVA including all continuous dependent variables relevant for this hypothesis. This analysis yielded a significant multivariate effect, $F(4, 168) = 3.27$, $p = .013$, $\eta_p^2 = .07$, indicating that there are systematic differences in the answers given by people with and without ASC.

Specifically, people with ASC spend more hours per week online, $F(1, 171) = 8.00$, $p = .005$, $\eta_p^2 = .045$ and spend more hours per week on CMC, $F(1, 171) = 5.66$, $p = .018$, $\eta_p^2 = .032$. People with ASC report to have made more acquaintances online than controls, $F(1, 171) = 10.31$, $p = .002$, $\eta_p^2 = .057$, but not to have made more friends online than controls, $F(1, 171) = 2.68$, $p = .104$. People with ASC further appreciate CMC much more than controls, $F(1, 171) = 16.59$, $p < .001$, $\eta_p^2 = .088$. Means are given in Table 2.

[insert Table 2 here]

People with ASC also answered significantly more often "yes" to the question whether they had found friends or acquaintances through CMC that they would not have known otherwise more often than controls (67.9% of ASCs vs. 42.3% of controls, $\chi^2(1) = 11.58$, $p < .001$). The MANOVA reported above showed that this difference only manifested for acquaintances and not friends. For both groups we found that the number of new acquaintances was higher than the number of new friends, which is in line with earlier findings that CMC mostly fosters weak ties (Ellison et al., 2007; Turner, Grube, & Meyers, 2001). However, ASCs and controls might have different operational definitions of "friends" and "acquaintances," a point we address in the discussion section.

We further investigated which specific channels are used more often by people with ASC than controls. Because usage of channels was not normally distributed, we conducted Mann-Whitney tests and found that the only channel used significantly more by people with ASC was formed by discussion sites, $U = 2562$, $z = -4.08$ (median and mean number of use are given in Table 3).

[insert Table 3 here]

However, this difference can be explained by the differences in recruitment for people with and without ASC, as recruitment for ASC relied more on requests posted on discussion sites than for controls. So, although people with ASC report spending more time using CMC, the difference cannot be found for most of the specific channels we investigated. It should be noted, however, that for most channels other than e-mail, ratings were very low, which severely restricts the possibility to find an effect.

In general, these results indicate that Internet use, CMC use, and number of acquaintances found online of respondents with ASC is greater than that of the control group and also that people with ASC value CMC more than controls. This is in line with our first hypothesis that people with ASC are especially attracted to CMC.

**Characteristics of CMC**

Having established that people with ASC make more use of CMC, we next investigated which aspects of CMC are seen as the most important advantages (again also comparing ASC to controls). We first analyzed the answers to the open questions. On average, people list 2.7 ($SD = 1.76$) advantages and 1.93 ($SD = 1.23$) disadvantages, with no differences between the two groups, $F(2, 181) = 1.40$, $p = .25$. We categorized the open answers into nine advantages and eight disadvantages. A MANOVA on the frequencies of the different advantages revealed that people with ASC and controls list different advantages $F(9, 174) = 7.40$, $p < .001$. For example, people with ASC most often list advantages which relate to the slower pace in CMC and controls most often list advantages which relate to the convenience of CMC (see Table 4 for the categories and the comparisons). There was no difference between the groups for the disadvantages, $F(8, 175) = 0.75$, $p = .65$.

**[insert Table 4 here]**

Second, we analyzed the ratings given to the 40 characteristics we formulated beforehand. Many of these features are named as beneficial in general (e.g. "Online I don't have to react instantly"), but some are seen as specifically advantageous for ASCs (e.g. "Online I don't have to pay attention to someone's facial expression"). Therefore, ASCs were expected to endorse a higher number of these statements, and endorse them more intensely.

We first performed factor analyses to identify the underlying constructs for the 40 characteristics to make further analyses more manageable. We followed recommendations from Costello & Osborne (2005) and used principal axis factors as extraction method because the scores on the individual items were not normally distributed. Based on the scree plot we decided how many factors to extract and we used direct oblimin rotation because we expect our factors to be correlated, given that they all measure aspects of CMC. We decided to exclude items with communalities lower than .40 because this indicates that the item is not related strong enough to other items in the analysis. After four rounds of excluding items three factors emerged which made theoretical sense, had at least five items, and had a good internal consistency, as described in the methods section. Factor loadings on the final three factors are presented in Table 5. We then performed a MANOVA with the scores on the three scales as dependent variables and group (ASC or control) as fixed factor. It yielded a significant multivariate effect, $F(3, 176) = 26.31$, $p < .001$, $\eta_p^2 = .31$. Moreover, it revealed significant differences between the ratings given by people with ASC and controls on all three subscales. As can be seen in Table 5, people with ASC perceive the timing of CMC to be more of an advantage than controls, they perceive the isolated communication context as more of an advantage and they perceive the relative ease to express oneself as more of an advantage.

**[insert Table 5 here]**

In fact, controls don't perceive the isolated communication context as an advantage at all, as their rating is below 50, whereas ratings of people with ASC are well above 50, and significantly so ($t(108) = 11.89$, $p < .001$).

When we compare the top-3 items of people with ASC and controls we see that both groups give high ratings to the fact that "Online I can communicate while being alone" and "Online I can choose at what time I want to communicate with others." For people with ASC the characteristic with the third highest rating is "Online I don't have to react instantly" which is fourth for controls. An interesting discrepancy emerges for controls' third rank, which is "I can combine online communication with other tasks", a characteristic which ranks 21 for people with ASC. It thus seems that there is quite some overlap in the characteristics of CMC which are seen as most advantageous, but there are also meaningful differences which are in line with the characteristics of ASC, because given their difficulty with multitasking, they do not perceive it as a strong advantage that CMC can be combined with other tasks.

In general, one can say that people with ASC ascribe more positive ratings to most characteristics of CMC than controls, as hypothesized. They see advantages in many characteristics of CMC, which is in line with their higher appreciation scores for CMC.

**CMC and well-being**

As a final step, we looked at several different outcome variables, to see whether CMC use in people with ASC affects satisfaction with different aspects of life. As a first step we investigated whether people with ASC and controls differ on the three satisfaction with life scales using a repeated measures ANOVA. We expected that people with ASC would have lower general life satisfaction, but we expected this difference to be less pronounced for their online social life, as the online social life should be easier for them to manage. There was a significant main effect for group, $F(1, 173) = 63.06$, $p < .001$, $\eta_p^2 = .27$, indicating that people with ASC report lower satisfaction than controls. There was also a significant main effect for type of outcome, $F(2, 346) = 18.85$, $p < .001$, $\eta_p^2 = .107$, indicating that scores for the three domains of satisfaction differ. Most importantly, and in line with our hypothesis, there was a significant interaction effect, $F(2, 346) = 30.29$, $p < .001$, $\eta_p^2 = .15$, indicating that the difference in satisfaction scores between the ASC group and the control group was not the same across the three domains. As can be seen in Figure 1, the difference in satisfaction between the two groups is much smaller when asked about their online social life as compared to their social life and life in general.

**[insert Figure 1 here]**

Furthermore, one-sample t-tests revealed that, whereas the mean satisfaction with their social life and their life is significantly lower than 4 (the mid-point of the scale), $t(109) = -4.27, p < .001$ and $t(109) = -4.21, p < .001$, satisfaction with their online social life for people with ASC is significantly *higher* than 4, $t(109) = 5.52, p < .001$.

To test the idea that CMC is positively related to well-being, we performed four regression analyses with intensity of CMC-use as a predictor variable for *satisfaction with online social life*, *satisfaction with social life*, *satisfaction with life*, and *loneliness*. In these analyses we used the score on the AQ to differentiate between people with and without ASC, because a continuous predictor yields a much more detailed picture and a statistically more robust result than a dichotomous predictor.[3]

The overall model for satisfaction with online social life was significant, $F(3, 167) = 2.80, p = .042, R^2 = .05$. The only significant predictor was CMC use, $\beta = .21, p = .011$. The more people make use of CMC, the more satisfied they are with their online social life. Given that there was no significant interaction with AQ, this relationship works in the same way for people with or without ASC. We should immediately note here that, given the correlational nature of our study, we cannot say anything about the direction of this effect. It can be the case that the more people make use of CMC, the more satisfied they get with their online social life, but also that people who are more satisfied with their online social life make more use of CMC. Also note that the AQ score was not a significant predictor of satisfaction with online social life, which is in line with the previous analysis, showing that satisfaction with one's online social life is independent of autistic traits.

The overall model for satisfaction with social life was significant, $F(3, 166) = 17.22$, $p < .001$, $R^2 = .24$. The only significant predictor was the AQ, $β = -.48$, $p < .001$. The higher someone scores on the AQ, the less satisfied they are with their social life. Given that there was no significant interaction with CMC use, this relationship is present in the same way for people regardless of how often they use CMC.

The overall model for satisfaction with life was significant, $F(3, 166) = 26.02$, $p < .001$, $R^2 = .32$. The AQ was a significant predictor, $β = -.56$, $p < .001$. The higher someone scores on the AQ, the less satisfied they are with their life. For this outcome, there was also a significant interaction between AQ and CMC use, $β = -.14$, $p = .047$. To interpret this interaction we standardized all scores and plotted the regression lines for high, mean, and low levels of the predictors (see Figure 2) and analyzed whether the simple slopes are significant for people who score high on the AQ (one SD above the mean) and people who score low on the AQ (one SD below the mean).

**[insert Figure 2 here]**

As can be seen, for people with low levels of AQ, use of CMC is slightly positively related to satisfaction with life, but this slope was not significant, $b = .124$, $t = 1.05$, $p = .30$. For people with high levels of AQ, CMC was significantly negatively related to satisfaction with life, $b = -.164$, $t = -2.12$, $p = .036$. Again, we can only speculate about the causal direction of this effect. Either people with higher scores on the AQ who spend more time with CMC become less satisfied with their lives or people with higher scores on the AQ who are not very satisfied with their lives start to use more CMC (maybe to find like-minded people) or there is a third variable which relates to both.

Finally, the overall model for loneliness was significant, $F(3, 166) = 23.59$, $p < .001$, $R^2 = .30$. The only significant predictor was the AQ, $β = -.54$, $p < .001$. The higher someone scores on the AQ, the lonelier they feel. Given that there was no significant interaction with CMC use, this relationship is present in the same way for people regardless of how often they use CMC.

In general, the analyses concerning the well-being scales show that people with ASC are less satisfied with their life, but that they are relatively satisfied with their online social life, indicating that they feel good about CMC. However, there was also a small negative effect in that people with ASC who use CMC frequently are actually less satisfied with their life. We will discuss the implications of these findings below.

**Discussion**

The current study is (as far as we know) the first study to compare CMC use in people with and without ASC. It contributes to our knowledge on CMC use and ASC in several different ways. First, we find that the frequency of use of CMC and the number of online contacts of high-functioning ASCs is greater than or equal to the control group. Second, we find that people with high-functioning ASC have more appreciation for textual, self-paced, communication aspects of CMC than controls. Third, people with ASC are relatively satisfied with their online social life; more so than with their social life and their life in general. They still do not reach the level of satisfaction of controls, but the difference is smaller than in the other aspects of life and on average, they are on the positive end on the scale. Finally, high levels of autistic traits, combined with high levels of CMC use are associated with low levels of satisfaction with life. Together, these results may have important implications for our view of CMC-use by people with ASC, as we discuss below.

**CMC use and well-being**

Our results suggest that people with high-functioning ASC who use CMC are at least as active online as other people that use CMC, and they report relatively high levels of satisfaction with their online social life. However, it is disputed whether such a high level of online activity is beneficial. Computers, games and the Internet are often regarded as harmful for the social development of children and adults, depriving them of time spent establishing social contacts using more conventional ways of interaction (Barak & Sadovsky, 2008; Bargh & McKenna, 2004; Finkenauer, et al., 2012; Kraut, et al., 1998; McKenna & Bargh, 2000; Sheeks & Birchmeier, 2007). Indeed, we find that individuals who score high on autistic traits and spend a lot of time using CMC are actually less satisfied with their lives. However, we also find that people with ASC make new acquaintances and friends online, even more so than controls, and that people with ASC are relatively satisfied with their *online social* lives. Our results imply that people with high-functioning ASC who use CMC show no lack of interest in social contact, and that they are able to build a satisfying online social life. This supports the view by Newton et al. (2009), that social-communicative impairments may not be an intrinsic defect in people with Autism Spectrum Conditions, but are compounded by conventional, rich, multi-modal, communication methods of face-to-face conversations. It seems that, in a different communications environment, autistic impairments in the conventional communicative domain may have less severe social repercussions. It may be that the increasing use of CMC in our modern society will make it easier for high-functioning ASCs to establish the social contacts that they are interested in.

**Characteristics of CMC**

Both the ASC group and the control group see advantages in CMC, but there are differences in the type of advantages they see. Both among the open questions and the predefined characteristics, the slower pacing appears to be the most important advantage of CMC for people with ASC. This may be caused by decreased demands on information processing capabilities, reducing the need for an immediate response. CMC also provides for more time to think to formulate an answer, with the option of rereading a message, enabling a more structured conversation form. Interestingly, the control group mentions the time independence as an advantage mainly because of convenience reasons (e.g. being able to answer at one's own time). For people with ASC the time independence is more important because it gives them more time to process the message.

The major difference between the two groups can be found in the characteristic 'absence of non-verbal communication.' This characteristic is listed as an advantage more often by people with ASC and listed as a disadvantage more often by controls. One might say that the fact that CMC offers reduced stimulus/single channel communication is evidence that CMC is a "poorer" kind of communication than face-to-face communication. However, for people with ASC it affords a mode of communication that suits them: no requirements for instant response/non-verbal communication, a single mode of communication, and a textual orientation.

All in all, people with ASC name and value advantages that help to mitigate their autistic impairments, while for controls aspects of convenience seem more relevant. Additionally, people with ASC are more positive in their appreciation of CMC, as evidenced by their scores on the statements regarding CMC qualities and on the appreciation scale.

**Limitations**

The current study is the first to compare CMC use of people with and without ASC. We found that, even though people with ASC tend to communicate less in general, their use of CMC is higher than that of controls. However, the use of a control group raises the question of whether the two groups are comparable, since we find significant differences on some demographic variables. Most of these differences can be explained because people with ASC tend to have lower levels of education given the same level of intelligence (Estes, Rivera, Bryan, Cali, & Dawson, 2011), partly because they have more difficulties with the transition from school to higher education (VanBergeijk, et al., 2008). The longer duration of CMC may be explained by the affinity many ASCs have for the Internet and computer based tools, or by the greater amount of time at hand (respondents with ASC have indicated being unemployed more often or having a disability allowance, possibly allowing them to spend more time online). Still, two main points for methodological improvement are (1) the recruitment method (participants self-selected into the study) and (2) the survey method (an online self-report survey was used). For example, both ASCs and controls have been recruited through online social networks, personal contacts, patient group flyers, discussion forums, and e-mail. Since participants in both groups self-selected into the study, we had no control over how this may have influenced their CMC use. For the ASC group recruitment emphasized online forums and patient groups, for the controls the personal and university network were the main recruitment channels.

A possible limitation of the scale used for the first hypothesis is inconsistent use of what respondents call "friend" or "acquaintance." We have allowed our subjects to use their own definitions of who they regard as "friend." Therefore, definitions of "friend" are likely to differ to some extent between subjects; we have not performed an analysis whether a bias existed. Also, we did not specify a time frame for this question, so if people with ASC started using CMC earlier in their life, this may explain why they gathered more acquaintances. We cannot rule out these alternative explanations, but still think that the friends and acquaintances acquired by ASC play a meaningful role in their lives, given the relative satisfaction with their online social life.

A possible further limitation lies in the use of the AQ. It meets with some resistance in the ASC community (e.g. AllieKat, 2011). Some participants stated in their comments that the questions are too stereotyped, and too much geared towards male autism. Also, the AQ is widely known, and freely available on the Internet. One respondent refused to fill out the AQ, because they were already acquainted with it and found it too biased towards autism stereotypes. Another disadvantage is that it is relatively easy to fill out the AQ in a way to avoid getting a high score. Still, we find scores that are very similar to those obtained in the original study by Baron-Cohen and colleagues (2001), when the AQ was not yet freely available. The ASC group's mean AQ in our study is 34.68 (SD = 7.88) compared to a mean AQ score of 35.8 *(SD* = 6.5) in the original study. Our control group scores 13.59 (SD 8.10), compared to 16.4 (SD 6.3) in the original study. We therefore do think that the AQ still serves as a good instrument to identify autistic traits.

As this study focused on people who use CMC, our conclusions do not necessarily generalize to people who do not use CMC. We found that intensity of CMC use by high-functioning ASCs was higher than in controls. However, our sample did not include people who do not use CMC. It could be that the proportion of non-users of CMC is larger among ASCs than among controls. It would be especially interesting to conduct further research among a more diverse population, including non-high-functioning ASCs, to study the use of CMC, for example controlling for the level of general intelligence in the ASC group and in the control group.

Another suggestion for future research may be to include the caregiver perspective of people with ASC. It may be that parents or other relatives are actually 'protecting' individuals with ASC by limiting or controlling 'friendships' by internet. It may be especially interesting to contrast whether and how CMC is appreciated and perceived by people with ASC and caregivers. Our data suggest that people with ASC benefit more from CMC than caregivers might think.

**Conclusion**

This work focused on high-functioning ASCs who already use CMC. The traditional view of autistic individuals is that they are loners, not interested in other people, and incapable of initiating or maintaining mutual relationships. From the results of our survey a different picture arises. The subjects in this study use the communication options afforded by networked computers at least as enthusiastically as controls, and are proficient and successful in their use. Our results indicate that the absence of the instant response/non-verbal communication requirement attracts high-functioning ASCs to get online, to make friends, and to have an online social life that is relatively satisfactory for them.

The point of view that computers and the Internet offer an alternative for creating meaningful social relationships for people with ASC, without consistent support, is not undisputed. We have only scratched the surface with this first study to compare CMC use in people with and without ASC, and much research still needs to be done to more fully understand the issues involving autism and online communication.

Benford, P., & Standen, P. (2009). The internet: a comfortable communication medium for people with Asperger syndrome (AS) and high functioning autism (HFA)? *Journal of Assistive Technologies, 3*(2), 44-53.

Burke, M., Kraut, R., & Williams, D. (2010). *Social use of computer-mediated communication by adults on the autism spectrum*. Paper presented at the Proceedings of the 2010 ACM conference on Computer supported cooperative work, Savannah, Georgia, USA. conference paper retrieved from http://portal.acm.org/citation.cfm?id=1718991#

Cheng, L., Kimberly, G., & Orlich, F. (2002). KidTalk: Online therapy for Asperger's Syndrome. [Technical Report MSR-TR-2002-08 Microsoft Research, Microsoft Corporation].

Costello, A. B., & Osborne, J. W. (2005). Best practices in exploratory factor analysis: Four recommendations for getting the most from your analysis. *Practical Assessment, Research & Evaluation, 10*(7).

Davidson, J. (2008). Autistic culture online: virtual communication and cultural expression on the spectrum. *Social & Cultural Geography, 9*(7), 791-806. doi: citeulike-article-id:3397138

Diener, E., Emmons, R. A., Larsen, R. J., & Griffin, S. (1985). The satisfaction with life scale. *Journal of Personality Assessment, 49*(1), 71-75. doi:10.1207/s15327752jpa4901_13

Ducheneaut, N., & Moore, R. J. (2005). More than just 'XP': learning social skills in massively multiplayer online games. *Interactive Technology and Smart Education, 2*(2), 89-100.

Ellison, N. B., Steinfield, C., & Lampe, C. (2007). The Benefits of Facebook "friends": Social Capital and College Students' Use of Online Social Network Sites. *Journal of Computer-Mediated Communication, 12*(4), 1143-1168.

Table 1

*Demographic variables for people with Autism Spectrum Conditions and controls. Values denote number and (percentage) of respondents, except for AQ, age and working hours per week which are mean (SD) values.*

| Variable | ASC | control |
|---|---|---|
| AQ total, 4-point scoring method (N=182) | 147.03 (19.09) | 99.39 (18.23) |
| AQ total, binary scoring method (N=182) | 34.68 (7.88) | 13.59 (8.10) |
| mean age (N=183) | 40.2 (12.3) | 40.5 (12.1) |
| sex (N=183) | | |
|    men | 62 (55.9%) | 28 (38.9%) |
|    women | 49 (44.1%) | 44 (61.1%) |
| relational status (N=181) | | |
|    single | 65 (59.6%) | 20 (27.8%) |
|    partner | 44 (40.4%) | 52 (72.2%) |
| living situation (N=183) | | |
|    independent | 94 (84.7%) | 69 (95.5%) |
|    non-independent (with parents, sheltered etc.) | 17 (15.3%) | 3 (4.2%) |
| main daytime occupation (N=181) | | |
|    paid employment | 42 (37.8%) | 42 (58.3%) |
|    retired | 2 (1.8%) | 2 (2.8%) |
|    student | 13 (11.7%) | 19 (26.4%) |
|    disability allowance | 32 (28.8%) | 2 (2.8%) |
|    unemployed, actively seeking | 6 (5.4%) | 4 (5.6%) |
|    not employed otherwise | 16 (14.4%) | 3 (4.2%) |
| working hours per week (N=78) | 32.00 (7.66) | 30.53 (7.12) |
| educational level (completed) (N=183) | | |
|    primary school | 4 (3.6%) | 0 (0.0%) |
|    lower vocational / intermediate secondary education | 12 (10.8%) | 3 (4.2%) |
|    intermediate vocational / higher secondary education | 39 (35.1%) | 12 (16.7%) |
| higher vocational education | 31 (27.9%) | 31 (43.1%) |
| university | 25 (22.5%) | 26 (36.1%) |
| highest educational level (including uncompleted) (N=183) | | |
| primary school | 1 (0.9%) | 0 (0.0%) |
| lower vocational / intermediate secondary education | 11 (9.9%) | 1 (1.4%) |
| intermediate vocational / higher secondary education | 21 (18.9%) | 10 (13.9%) |
| higher vocational education | 35 (31.5%) | 29 (40.3%) |
| university | 43 (38.7%) | 32 (44.4%) |

Table 2

*Means and standard deviations for different indices of Internet use for people with ASC and Controls*

|  | ASC | | Controls | |
|---|---|---|---|---|
|  | *M* | *SD* | *M* | *SD* |
| Hours of Internet use per week | 23.20 | 16.07 | 16.46 | 13.96 |
| Hours spent on CMC per week | 13.95 | 14.41 | 9.01 | 11.25 |
| Number of new friends through Internet | 1.69 | 2.61 | 1.07 | 2.04 |
| Number of new acquaintances through Internet | 4.71 | 4.32 | 2.63 | 3.87 |
| Appreciation of CMC | 5.08 | 1.47 | 4.22 | 1.16 |

Table 3

*Median and mean number of uses of different CMC channels for people with ASC and Controls*

|  | ASC | | Controls | |
|---|---|---|---|---|
|  | *Mdn* | *M* | *Mdn* | *M* |
| E-mail | 5 | 4.40 | 5 | 4.33 |
| Twitter | 0 | 0.18 | 0 | 0.62 |
| Text chat | 0 | 1.25 | 0 | 1.28 |
| Audiovisual chat | 0 | 0.28 | 0 | 0.41 |
| Social network sites | 0 | 1.30 | 1 | 1.67 |
| Professional network sites | 0 | 0.67 | 0 | 0.65 |
| Discussion sites and forums | 1 | 1.89 | 0 | 0.64 |
| Dating sites | 0 | 0.17 | 0 | 0.01 |
| Games | 0 | 0.66 | 0 | 0.30 |

Table 4

*Answer categories for advantages and disadvantages of CMC as mentioned by ASCs and controls with mean frequencies (SD), and the comparison*

|  | ASC | Controls | F-value |
|---|---|---|---|
| Advantages |  |  |  |
|    Time independence: Pacing | 0.82 (0.90) | 0.35 (0.66) | 14.58*** |
|    Time independence: Timing | 0.29 (0.56) | 0.62 (0.64) | 13.30*** |
|    No co-presence: Absence of non-verbal communication | 0.25 (0.49) | 0.07 (0.26) | 7.87** |
|    No co-presence: Anonymity / invisibility | 0.11 (0.31) | 0.04 (0.20) | 2.39 |
|    Less sensory overload | 0.19 (0.45) | 0.06 (0.23) | 4.95* |
|    Textual form | 0.37 (0.62) | 0.17 (0.48) | 5.59* |
|    Enhanced contact, social skills | 0.22 (0.59) | 0.23 (0.74) | 0.00 |
|    Convenience | 0.30 (0.65) | 1.01 (1.02) | 33.43*** |
|    Decreased stress | 0.14 (0.38) | 0.01 (0.12) | 7.72** |
| Disadvantages |  |  |  |
|    Time independence: Too slow | 0.20 (0.45) | 0.23 (0.48) |  |
|    No co-presence: Absence of non-verbal communication | 0.35 (0.55) | 0.41 (0.52) |  |
|    No co-presence: Lack of direct contact | 0.65 (2.05) | 0.44 (0.69) |  |
|    Usage intensity | 0.04 (0.19) | 0.11 (0.36) |  |
|    Textual form | 0.19 (0.41) | 0.18 (0.43) |  |
|    Technique-related issues | 0.20 (0.50) | 0.24 (0.64) |  |
|    Friction | 0.27 (0.45) | 0.32 (0.50) |  |
|    Miscellaneous | 0.05 (0.23) | 0.06 (0.23) |  |

\* $p < .05$; \*\* $p < .01$, \*\*\* $p < .001$.

Table 5

*Scores on statements regarding CMC qualities. Values denote Mean (SD) for each respondent group, and t-value.*

|  | ASC | | Controls | | Comparison / Factor loading |
|---|---|---|---|---|---|
|  | M | SD | M | SD |  |
| **Scale: Time independence** | 76.48 | 12.69 | 65.93 | 9.34 | $F(1, 178) = 36.21^{***}$ |
| Online I can have a conversation in a quiet environment | 82.49 | 18.27 | 70.85 | 17.70 | .742 |
| Online I don't have to react instantly | 82.61 | 18.91 | 71.03 | 16.96 | .658 |
| Online I can communicate while being alone | 85.39 | 18.81 | 74.94 | 17.96 | .656 |
| Online I can take the time to formulate correctly what I want to say | 81.99 | 15.52 | 65.24 | 15.37 | .642 |
| Online I can choose at what time I want to communicate with others | 83.84 | 19.01 | 85.41 | 12.82 | .633 |
| Online I can communicate while being in my own familiar surroundings | 79.16 | 19.27 | 64.48 | 18.86 | .611 |
| Online I can take the time to process what the other person says | 78.72 | 17.08 | 62.45 | 14.57 | .564 |
| Online I can write my responses when I want, regardless of whether my conversation partner is online simultaneously | 80.42 | 21.37 | 70.42 | 17.58 | .546 |
| Online you can get to the point immediately | 70.39 | 20.10 | 63.61 | 15.62 | .511 |
| Online I can better express myself | 73.65 | 21.08 | 55.96 | 14.16 | .491 |
| Online there's less social chit-chat | 72.55 | 24.84 | 56.30 | 16.58 | .380 |
| Online I can directly contact people that I would not be able to reach otherwise | 74.28 | 19.90 | 68.23 | 18.00 | .343 |
| **Scale: No co-presence** | 67.49 | 15.35 | 49.58 | 9.66 | $F(1, 178) = 76.77^{***}$ |
| Online I don't have to watch my facial expression | 69.41 | 21.11 | 49.58 | 15.69 | .792 |
| Online I don't have to make eye contact | 72.07 | 22.94 | 47.58 | 14.33 | .714 |
| Online I don't have to pay attention to the other's facial expression | 69.11 | 23.76 | 48.27 | 14.29 | .660 |
| Online nobody can see me | 65.10 | 21.27 | 51.66 | 12.77 | .629 |
| Online I don't have to pay attention to the way someone sounds | 61.62 | 24.87 | 48.27 | 14.30 | .600 |
| Online people express themselves more clearly | 61.33 | 17.96 | 49.49 | 9.52 | .438 |
| Online I only have to pay attention to what is written | 73.75 | 20.71 | 51.94 | 13.40 | .432 |
| Scale: Relative ease to express oneself | 61.39 | 14.33 | 53.90 | 9.13 | $F(1, 178) = 15.32^{***}$ |
| Through my online experience I can have a real-life conversation more easily | 56.42 | 17.82 | 52.00 | 10.35 | .625 |
| Online I can talk in a more personal way with others | 61.20 | 20.45 | 51.82 | 12.31 | .618 |
| Online I can more easily bring up a difficult subject | 64.74 | 18.88 | 54.80 | 12.59 | .581 |
| Online I can disclose more about myself | 59.17 | 21.00 | 53.41 | 13.16 | .468 |
| Through my online experience I can have an online conversation more easily | 65.40 | 22.80 | 57.46 | 16.58 | .431 |
| Items not in a scale: | | | | | |
| Online I can remain anonymous | 59.11 | 23.77 | 54.10 | 17.74 | |
| I can combine online communication with other tasks | 65.17 | 24.53 | 74.04 | 18.53 | |
| I receive a lot of emails | 52.69 | 21.11 | 53.20 | 17.65 | |
| Online I can choose to contact someone based on their profile | 59.17 | 19.99 | 53.45 | 11.19 | |
| Keeping up with all my contacts takes much time | 41.00 | 17.18 | 44.01 | 13.67 | |
| Online people are often sloppy in their writing | 36.55 | 18.96 | 40.96 | 17.75 | |
| Online people can be rude or insulting | 35.09 | 18.56 | 39.96 | 19.47 | |
| Online many people can take part in a conversation or discussion simultaneously | 45.83 | 24.88 | 57.46 | 18.04 | |
| In chat programs conversations are often very high-paced | 38.46 | 22.96 | 48.21 | 15.36 | |

| | | | | |
|---|---|---|---|---|
| I can illustrate my remarks with documents or links | 65.70 | 18.81 | 61.24 | 16.86 |
| I don't know if I can trust my online conversation partners | 36.55 | 19.04 | 41.52 | 12.78 |
| Online nobody can see that I have a disability | 61.75 | 23.04 | 49.90 | 10.76 |
| What I write online has permanence | 55.30 | 28.31 | 54.63 | 24.87 |
| An online conversation is usually slower than a real-life conversation | 63.89 | 24.02 | 48.61 | 15.53 |
| Online I don't have to pay attention to the way I come across | 66.83 | 22.73 | 53.70 | 14.53 |
| During online conversations I can always read back what exactly has been written | 75.81 | 20.95 | 65.42 | 17.39 |

*** $p < .001$.

Computer-mediated communication in adults with high-functioning ASC                38Figure Captions

*Figure 1*. Satisfaction scores for different aspects of life for people with ASC and controls.

*Figure 2*. Relationship between CMC use and Satisfaction with Life at different AQ scores. Low represents -1 SD, Mean represents the mean value, and High represents +1 SD for both CMC use and AQ score.

Footnotes

[1] Of course these websites are also of interest and probably also consulted by family members of people with ASC. However, given that we were able to easily recruit ASC participants via these websites indicates that people with ASC also frequently use these sites.

[2] To be sure, we also inspected the scores on the AQ of these individuals. These scores are well above the scores of the control group (22, 27, 28, 34, 41).

[3] Using group as a dummy variable yielded almost the same result, with the exception that the interaction between group and CMC use on Satisfaction with life was only marginally significant, $\beta = -.253$, $p = .073$.